# Investigation of the microstructure of the fine-grained YPO$_4$:Gd ceramics with xenotime structure after Xe irradiation

Dmitriy A. Mikhaylov [1,*], Ekaterina A. Potanina [1], Aleksey V. Nokhrin [1,*], Albina I. Orlova [1], Pavel A. Yunin [1,2], Nikita V. Sakharov [1], Maksim S. Boldin [1], Oleg A. Belkin [1], Vladimir A. Skuratov [3,4,5], Askar T. Isatov [3,6,7], Vladimir N. Chuvil'deev [1], Nataliya Y. Tabachkova [8,9]

[1] Materials Science Department, Physical and Technical Research Institute, Lobachevsky State University of Nizhny Novgorod, 603022, Nizhny Novgorod, Russia; mikhaylov@ichem.unn.ru (D.M.), potanina@nifti.unn.ru (E.P.), nokhrin@nifti.unn.ru (A.N.), albina.orlova@gmail.com (A.O.), nvsaharov@nifti.unn.ru (N.S.), boldin@nifti.unn.ru (M.B.), chuvildeev@nifti.unn.ru (V.C.)

[2] Laboratory of Diagnostics of Radiation Defects in Solid State Nanostructure, Institute for Physics of Microstructure, Russian Academy of Science, 603950, Nizhniy Novgorod, Russia; yunin@ipmras.ru (P.Y.)

[3] G.N. Flerov Laboratory of Nuclear Reactions, Joint Institute of Nuclear Research, 141980, Dubna, Russia; skuratov@jinr.ru (V.S.), issatov@jinr.ru (A.I.)

[4] Institute of Nuclear Physics and Engineering, National Research Nuclear University MEPhI (Moscow Engineering Physics Institute), 115409, Moscow, Russia; skuratov@jinr.ru (V.S.)

[5] Department of Nuclear Physics, Dubna State University, 181982, Dubna, Russia; skuratov@jinr.ru (V.S.)

[6] Gumilov Eurasian National University, 010000, Nur-Sultan, Kazakhstan; issatov@jinr.ru (A.I.)

[7] The Institute of Nuclear Physics, 050032, Almaty, Kazakhstan; issatov@jinr.ru (A.I.)

[8] Center Collective Use "Materials Science and Metallurgy", National University of Science and Technology "MISIS", 119991, Moscow, Russia; ntabachkova@misis.ru (N.T.)

[9] Laboratory "FIANIT", Laser Materials and Technology Research Center, A.M. Prokhorov General Physics Institute, Russian Academy of Sciences, 119991, Moscow, Russia; ntabachkova@gmail.com (N.T.)

* Correspondence: mikhaylov@ichem.unn.ru (D.M.), nokhrin@nifti.unn.ru (A.N.)

## 1. Introduction

Compounds with the xenotime structure are among the potential matrices for actinide immobilization [1–3]. The xenotime structure consists of the PO$_4$ tetrahedrons and of the YO$_8$ polyhedrons, its crystallizes in the tetragonal syngony (space grout $I4_1/amd$) [4]. The structure is characterized by a wide set of isomorphic forms and may incorporate lanthanides (from Tb to Lu), rare earth elements Y and Sc [4], Pu, Cm, Np [5–7], Gd, Dy, Er, Yb [8,9], Th, U [2,9,10]. Due to the presence of Th and U in the composition, the natural xenotime compounds can be exposed to a radiation dose of up to $(1.4–14) \cdot 10^{16}$ $\alpha$/mg [10]. This characterizes indirectly the xenotime structure as the one having a good resistance to self-irradiation. In addition, the compounds with the xenotime structure have a high melting point [11,12] and are stable under hydrolytic conditions [11,13].

In [14], the amorphization of synthetic xenotime samples under irradiation with Kr$^{2+}$ ions (800 keV) was investigated as a function of temperature (20–600 K). It has been shown that at elevated temperatures, xenotime is more susceptible to radiation exposure. Synthetic LuPO$_4$ containing 1.0 wt.% $^{244}$Cm accumulated a dose of $5 \cdot 10^{16}$ $\alpha$/mg after 18 years of exposure and remained in a highly crystalline state, but with the formation of some defect regions in the samples, which could contain accumulated radiogenic helium [10]. The implantation of the Au ions (2 MeV, $1 \cdot 10^{14}$, $5 \cdot 10^{14}$, and $1 \cdot 10^{15}$ cm$^{-2}$) caused the structural damage in phosphates La$_{1-x}$Yb$_x$PO$_4$ (x = 0.7, 1.0) [15]. After irradiation with the maximum dose, the samples recrystallized partially. After annealing at 300 °C, these ones recovered completely. In [16], the YPO$_4$ ceramics were implanted with Au$^-$ ions with different energies (35, 22, 14, and 7 MeV) and fluences ($1.6 \cdot 10^{13}$ – $6.5 \cdot 10^{13}$ ions/cm$^2$). The authors supposed the heterogenous damaged layer observed in the YPO$_4$ ceramics to be a result of the epitaxial annealing occurring at the crystal-amorphous phase interface. In [17], when studying the irradiation of ErPO$_4$ ceramics with Au and He ions, He ions were found to prevent the amorphization of the sample partially at the simultaneous irradiation.

The ceramics based on the compounds with the xenotime structure are obtained usually by the pre-pressing of the powders followed by conventional sintering at the temperatures of ~ 1300–1600 °C; the sintering times are 1–5 h. This allows obtaining the ceramics with the relative density of ~90% [18,19]. To ensure an increased radiation and hydrolytic resistance of xenotime-structured ceramics, it is necessary to provide higher density values. In [20], YPO$_4$ ceramic with the relative density of 98% was obtained by a long sintering at 1600 °C during 10 h. The authors of [21] suggested that the sintering of the YPO$_4$ ceramics is difficult probably owing to the features of the particle morphology – an anisotropic crystal growth and the formation of the needle-shaped particles take place when heating up to 1400 °C.

Spark Plasma Sintering (SPS) is a promising methods for obtaining the ceramics for the immobilization of high level waste (HLW) [22–26]. SPS is a new method of high-speed (up to 2500 °C/min) hot pressing [22]. A possibility to achieve an increased density of the ceramics in short sintering times is an important advantage of the SPS technology [24–28]. This plays an important role in the handling of HLW. Mineral-like ceramics obtained by SPS have high relative density, hydrolytic, and radiation resistance [25,31–35]. The compaction kinetics of the nano- and submicron particles in SPS is determined by the intensity of the grain boundary diffusion [33-38]. The grain boundary diffusion coefficient for the fine-grained ceramics at low heating temperatures is known to be several orders of magnitude greater than the one inside the crystal lattice diffusion coefficient [39,40]. It allows obtaining the ceramic specimens with high relative density by SPS at low temperatures and increased heating rates [25,27–35].

The goal of the present work was to obtain Y$_{0.95}$Gd$_{0.05}$PO$_4$ (YPO$_4$:Gd) high-density ceramics with the xenotime structure by SPS and to study the microstructure and properties of the ones. In particular, we have studied the effect of irradiation on the microstructure and properties of the YPO$_4$:Gd ceramics including the formation of the amorphous phase. Gd$^{3+}$ was chosen as a simulator of Cm$^{3+}$ due to the similarity of their electronic configurations and ionic radii, and, therefore, of the chemical and physical properties. The irradiation with heavy ions in a charged particle accelerator created harsh conditions for the radiation exposure, which provides an accelerated assessment of the radiation resistance of the material.

## 2. Materials and Methods

Phosphate samples Y$_{0.95}$Gd$_{0.05}$PO$_4$ were obtained by the sol-gel method. Crystalline Y(NO$_3$)$_3$·6H$_2$O, Gd$_2$O$_3$ dissolved in an excess of a weakly acidic solution of HNO$_3$ (pH = 4 – 5) were used as the initial reagents. Ammonium dihydrogen phosphate NH$_4$H$_2$PO$_4$ as was used as a precipitant. The 1M solution of NH$_4$H$_2$PO$_4$ was added dropwise with continuous stirring. The resulting gel was stirred further vigorously for 5 min until complete homogenization and dried at 90 °C for a day. The dry residue was dispersed in an agate mortar. The resulting solid mixture was heated sequentially up to 600, 700, 800, and 900 °C for 5 h at each stage without any intermediate dispersion.

YPO$_4$:Gd ceramic samples with a diameter of 10 mm in diameter and 3 mm in height were obtained by the SPS method using Dr. Sinter® model SPS-625 equipment (NJS Co., Ltd., Tokyo, Japan). Sintering was carried out in vacuum (5-6 Pa), in graphite molds. To improve the contact of the powder sample with the graphite mold and to compensate for the difference in the thermal expansion coefficients of graphite and the ceramics, a graphite paper was used, which was placed inside the mold. The value of the uniaxial pressure applied was P = 70 MPa. The pressure was applied simultaneously with the start of heating. Two-stage heating was used: Stage I - heating up to 600 °C with the heating rate $V_h$ = 100 °C/min; Stage II - heating with $V_h$ = 50 °C/min up to the sintering temperature $T_s$ (Fig. 1a). The holding time at the sintering temperature $T_s$ was $t_s$ = 2 min. Total time of the sintering process was approximately 18 min. The sample was cooled down together with the Dr. Sinter® model SPS-625 setup.

The temperature was measured using CHINO® IR-AHS2 infrared pyrometer (Chino Corporation, Tokyo, Japan) focused on the surface of the graphite mold. Based on previous studies and on the comparison of the data measured with the optical pyrometer, (curve T1 in Fig. 1a) and of an additional thermocouple attached to the sample surface, the values of T1 were recalculated to the actual sample temperature (T2) using the empirical equation: T2 = 1.1686·T1- 43.416 (Fig. 1a).

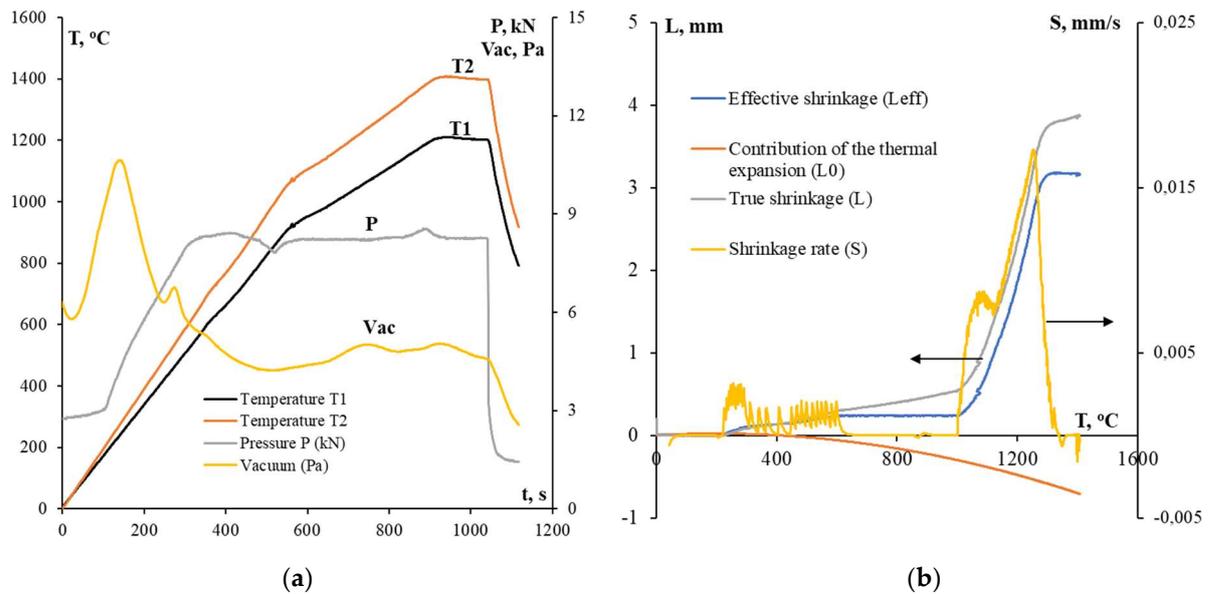

**Figure 1.** SPS diagrams of YPO$_4$:Gd ceramics: (a) dependences of temperature (T), applied pressure (P) and vacuum pressure (Vac) on the time of the SPS process; (b) dependence of the shrinkage on the heating temperature.

In the SPS process, the dependencies of the effective shrinkage ($L_{eff}$) on the sintering time and on the heating temperature were recorded. To take into account the contribution of the thermal expansion ($L_0$), special experiments were carried out on heating the empty molds (see [41]). The true shrinkage value was calculated as $L = L_{eff} - L_0$ (Fig. 1b). Using the temperature dependence L(T) in the linear approximation, the temperature dependence of the shrinkage rate was calculated: $S = \Delta L/\Delta t$.

After sintering, the sample surfaces were contaminated by residual graphite paper. To remove the residual graphite, the samples were annealed in EKPS-10 (Smolensk SKTB SPU, JSC., Smolensk, Russia) air furnace at 750 °C, 1 h. The samples were heated up and cooled down together with the furnace.

X-Ray Diffraction (XRD) phase analysis of the powders and ceramics was carried out using Shimadzu® LabX™ XRD-6100 diffractometer (Shimadzu Co., Kyoto, Japan). The semiquantitative phase analysis was performed using PhasanX® v.2.0 software.

The Grazing Incidence XRD (GIXRD) phase analysis of irradiated ceramics was performed using Bruker® D8 Discover™ X-ray diffractometer (Bruker Co., Billerica, US). An X-ray tube with a Cu cathode (CuK$_\alpha$ radiation) was used. The investigations were carried out in the parallel beam geometry with a parabolic Göbel mirror, a round collimator of 1 mm in diameter on the primary, and a 0.2° Soller slit in front of the detector. In every series of the GIXRD experiments, the angle of incidence of the primary beam $\alpha$ onto the specimen varied from 2° to 10°. In each experiment, scanning was performed by the detector in the angle 2θ. The XRD curves were recorded by scanning the detector in the range of 2θ = 25°–28° corresponding to the position of the 200 XRD peak of YPO$_4$.

The microstructure of the powders and ceramics was investigated using Jeol® JSM-6490 Scanning Electron Microscope (SEM, Jeol® Ltd., Tokyo, Japan) with Oxford Instruments® INCA 350 Energy Dispersion Spectroscopy (EDS) microanalyzer (Oxford Instruments® pls., Abingdon, UK) and Jeol® JEM-2100 Transmission Electron Microscope (TEM, Jeol® Ltd., Tokyo, Japan). The mean particle size (R) and grain size (d) was calculated by section method using GoodGrains® 2.0 software (UNN, Nizhny Novgorod, Russia). The microstructure of the ceramic surface layers after irradiation was studied using a Leica® IM DRM metallographic optical microscope (Leica Microsystems, Wetzlar, Germany).

The density of the obtained ceramics was measured by hydrostatic weighing in distilled water using Sartorius® CPA 225D balance (Sartorius AG, Göttingen, Germany). The uncertainty of the density measurement was ± 0.001 g/cm$^3$. The theoretical density of the ceramics ($\rho_{th}$) was calculated on the base of the XRD investigations.

The radiation resistance of the ceramics was evaluated using high energy (167 MeV) Xe$^{+26}$ ion irradiation at IC-100 FLNR JINR cyclotron (Joint Institute of Nuclear Research, Dubna, Russia). The samples were irradiated at the temperatures of 23-27 °C with the fluences (F) $1·10^{12} – 3·10^{13}$ cm$^{-2}$. The average ion flux was ≈$2·10^9$ cm$^{-2}$·s$^{-1}$ to avoid any significant heating of the targets. The temperature of the targets during the irradiation did not exceed 30 °C. The uniform distribution of the ion flux over the irradiated target surface was achieved by ion beam scanning. The accuracy of the ion flux and fluence measurements was 15%.

## 3. Results and Discussion

According to the XRD data (curve 1 in Fig. 2), the synthesized $Y_{0.95}Gd_{0.05}PO_4$ phosphate powder was monophasic and was identical to the analogue (ICDD PDF 83-0658, sp.gr. $I4_1/amd$, $a = b = 6.8905 \pm 0.0003$ Å, $c = 6.0227 \pm 0.0004$ Å, $\alpha = \beta = \gamma = 90°$). The theoretical density of the $YPO_4$:Gd compound calculated from the analysis of the results of the XRD studies is $\varrho_{th} = 4.349 \pm 0.001$ g/cm³. According to the XRD phase analysis results, there were no impurity phases in the powders.

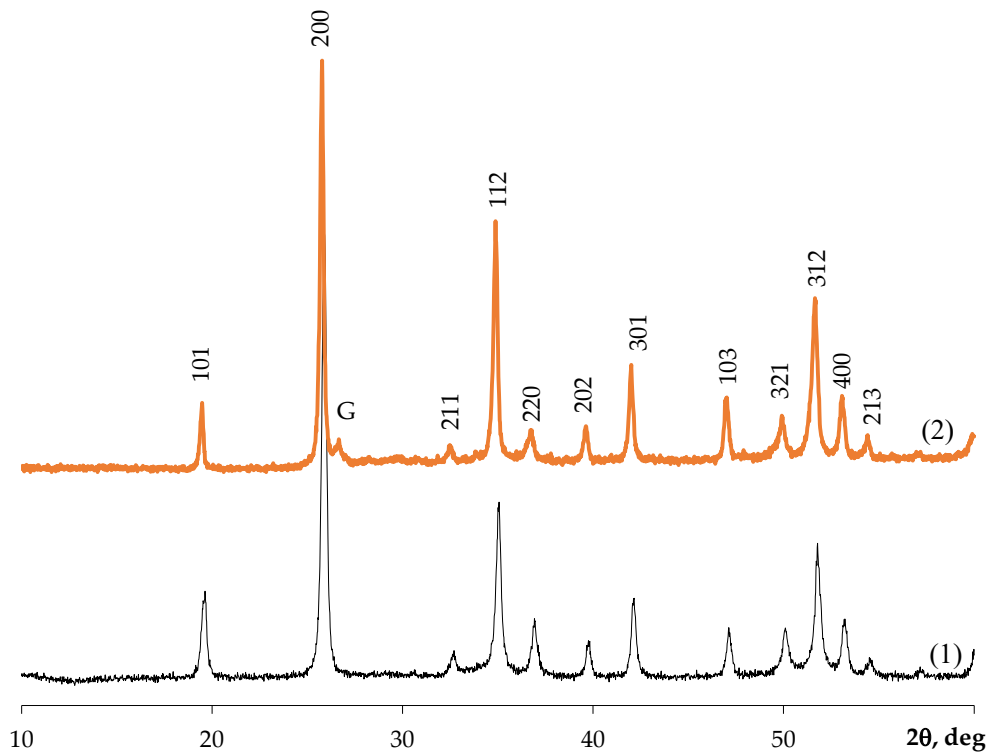

**Figure 2.** XRD data for YPO$_4$:Gd powder after annealing at 900 °C (5 h) (1) and for the SPS ceramic (2). G – symbol marks the reflection from graphite.

The SEM results show the synthesized powders to form large agglomerates with the sizes ranging from 10 to 50 μm (Fig. 3a). The agglomerates consist of the nanoparticles packed closely to each other (Fig. 3b). The conclusion that the agglomerates consist of individual nanoparticles was confirmed by the results of the TEM studies (Fig. 4). No large needle-shaped particles were observed in the powders (see [21]). Probably, it is related to the synthesis of the powders at lower temperatures in the present work than in [21].

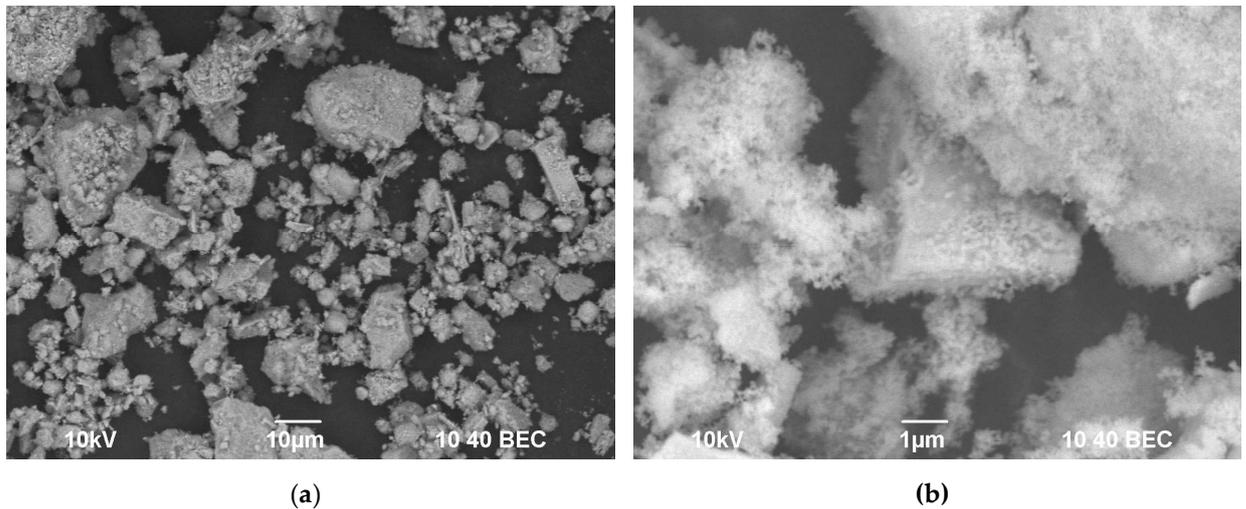

**Figure 3.** SEM images of YPO$_4$:Gd powder (**a**, **b**). Figs. 3a and 3b present various photographs of the same powder at different magnifications.

One can see in Fig. 4a that the YPO$_4$:Gd powder consisted of agglomerates, which, in turn, consisted of nanoparticles. The sizes of the agglomerates varied in a broad range – from several microns to several tens microns. The closed pores were observed in some particles. The particles had a crystalline structure. It was confirmed by the diffraction patterns measured from a selected area and the high resolution image of the particles (Fig. 4c, d). The positions of the diffraction maxima in the electron diffraction pattern (Fig. 4b) matches to the YPO$_4$ phase. The diffusion of the rings in the electron diffraction pattern also evidences the fine dispersed structures of the powder. The large particles in the powders were elongated mainly. The shapes of smaller particles were close to the spherical ones. In the statistical analysis of the size distribution of the particles, the shapes of the elongated particles were approximated by ellipses. When plotting the size distributions of the particles (Fig. 4c), both bigger and smaller axes of the ellipses were taken into account. The particle sizes ranged from 20 to 90 nm. The maximum of the particle size distribution corresponds to 50 nm (Fig. 4c).

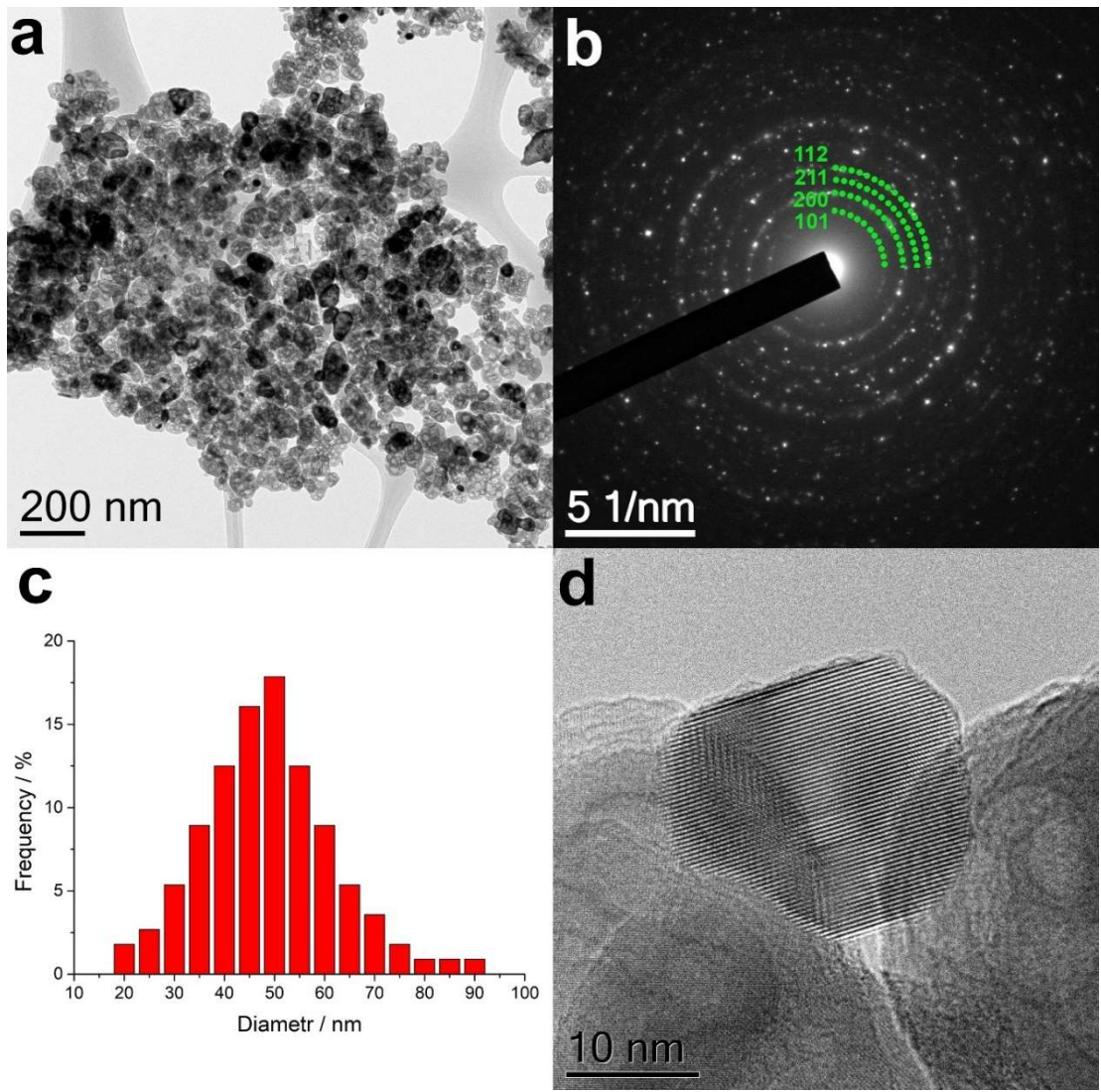

**Figure 4.** Results of TEM studies of YPO$_4$:Gd phosphate powder: (a) bright field image of the powder, (b) electron diffraction pattern, (c) particle size distribution histogram, (d) the crystal structure of the particles.

Seven identical ceramic samples were fabricated by SPS. The relative density of the sintered ceramic samples was 98.67 ± 0.18 %. The ceramic samples had no external damage and macrodefects.

Fig. 1a shows the sintering diagrams "time t – temperature T – pressure P – vacuum pressure Vac" for the YPO$_4$:Gd ceramics. Note that the vacuum pressure during heating remains approximately constant (4.5–4.8 Pa). This indicates the absence of decomposition of the YPO$_4$:Gd phase during SPS as well as the absence of significant dissociation of elements from the ceramic sample surfaces. Fig. 1b presents the temperature dependencies of the shrinkage (L) and of the shrinkage rate (S) for YPO$_4$:Gd powders. The dependence L(T) has usual three-stage character - stage I with a low compaction intensity (T < 1000 °C), stage II of intensive compaction of powders in the temperature range from 1000 to 1300 °C, and, finally, stage III where the intensity of powder shrinkage becomes small again (T > 1300 °C).

It is interesting to note that two maxima are visible clearly in the temperature dependence of the shrinkage rate: at ~1100 °C (the maximum shrinkage rate reaches $S_{max}$ ~ 8·10$^{-3}$ mm/s) and at ~1250 °C ($S_{max}$ ~ 17·10$^{-3}$ mm/s). In our opinion, the two-stage character of the dependence S(T) was related to the microstructure of YPO$_4$:Gd powders (Figs. 3, 4). At the first stage, the YPO$_4$:Gd phosphate nanoparticles are sintered inside the agglomerates. At the second stage, the agglomerates are sintered to each other.

It is important to note that the sintering temperature for the ceramics from the powder with the xenotime structure by SPS was lower than the one in other processes reported in the literature [18–20,25], and the duration of the SPS process was much shorter. The results obtained indicate a high workability and efficiency of SPS for producing the ceramics from polycrystalline inorganic powders (for example, the YPO$_4$:Gd with compound with the xenotime structure studied in the present work). A significant decrease in the temperature and duration of the sintering process by SPS makes it possible to prevent the growth of crystallites and obtain a material with a high relative density, low porosity, and a denser microstructure.

According to the XRD results, the phase composition of the ceramic samples was similar to the one of the initial powders (curve 2 in Fig. 2). There is a low intensive graphite peak in the XRD pattern of the ceramics. The presence of this peak was associated with the sintering in the graphite mold.

The SEM images of the fractures of ceramic samples are shown in Fig. 5. The micrographs show the ceramic to have a highly dense fine-grained microstructure. The majority of the ceramic grains were from 5 to 15 μm in size (Fig. 5a). Also, there were some parts with a fine-grained microstructure where the grain sizes were 1-2 μm. There were some isolated pores within these fine-grained structures. In some parts of the samples, large grains were present. In our opinion, this is a consequence of the presence of large particles and agglomerates in the original powders, not of the accelerated grain growth during sintering by SPS.

The results of the metallographic studies presented in Fig. 5 also indicate the formation of a high-density fine-grained microstructure with the grain size of ~15–30 μm. Thus, rather intensive grain growth took place during the SPS. It is interesting to note that the grain sizes in the ceramics (~10–20 μm, see Fig. 5a) were close to the sizes of the agglomerates in the initial powder (see Fig. 3a). The fractographic analysis of the fractures revealed the parts consisting of submicron particles in some samples. The sizes of these parts were also close to the sizes of agglomerates and to those of the grains (Fig. 3a, 5b). In our opinion, the results obtained also indicate indirectly the sintering process to have a two-stage character (see above). **There were some large elongated grains in the central parts of the samples (Fig. 5c, d).** Note that the SPS temperature for the YPO$_4$:Gd ceramics in the present work corresponded to the synthesis temperature for the YPO$_4$ powders in [21], which the anisotropic particle growth was reported in. It allows suggesting the large elongated grains in the YPO$_4$:Gd ceramics made by SPS to originate also from the anisotropic particle growth along selected crystallographic direction.

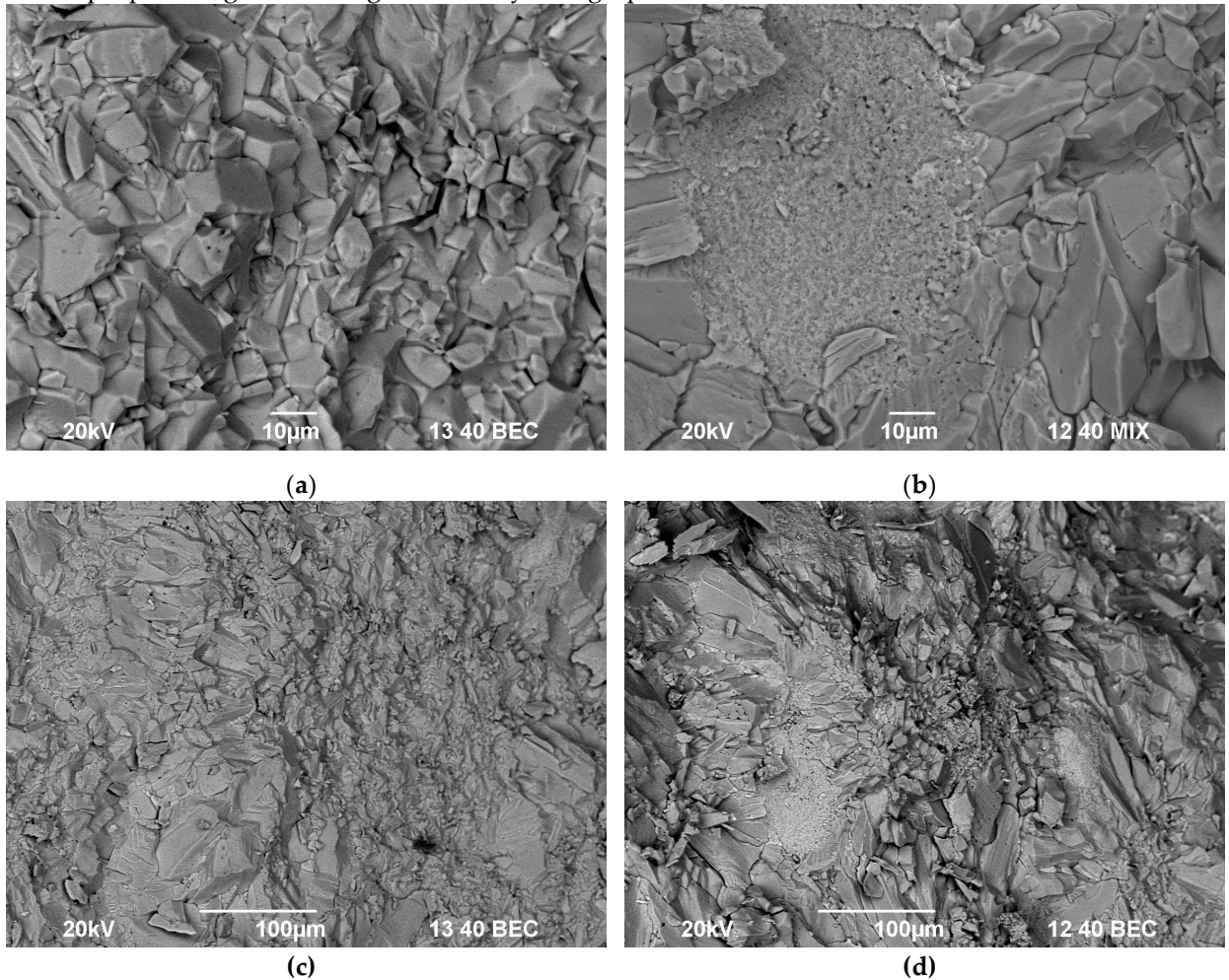

**Figure 5.** Microstructure of the YPO$_4$:Gd ceramics obtained by SPS. (a) metallography; (b) SEM of the sample fracture surface.

The ceramic samples were irradiated with $Xe^{+26}$ ions with the fluences F = $1\times10^{12}$, $3\times10^{12}$, $7\times10^{12}$, $1\times10^{13}$, and $3\times10^{13}$ cm$^{-2}$. The XRD results for the irradiated samples are presented in Fig. 6a. One can see in Figs. 6a and 6b that the increasing of the fluence leads to the decrease in the relative intensity of the 200 $YPO_4$:Gd peak. It is interesting to note that the peak from graphite in the irradiated ceramics became much more visible than in the non-irradiated one.

The diffraction pattern observed indicates the absence of complete amorphization of the surface layer of the ceramic sample: the XRD peaks corresponding to the $YPO_4$:Gd phase are seen quite clearly even after the irradiation with the fluence of $3\times10^{13}$ cm$^{-2}$ (Fig. 6a). We emphasize that the broadening and decrease in the intensity of the XRD peaks (at the diffraction angles $2\theta$ = 25–28°) indicate an increased fraction of the amorphous phase on the sample surfaces. The dependencies of the 200 peak intensity on the fluence are shown in Fig. 6b, c. As one can see in Fig. 6c, an exponential decay of the 200 XRD peak intensity with increasing fluence was observed.

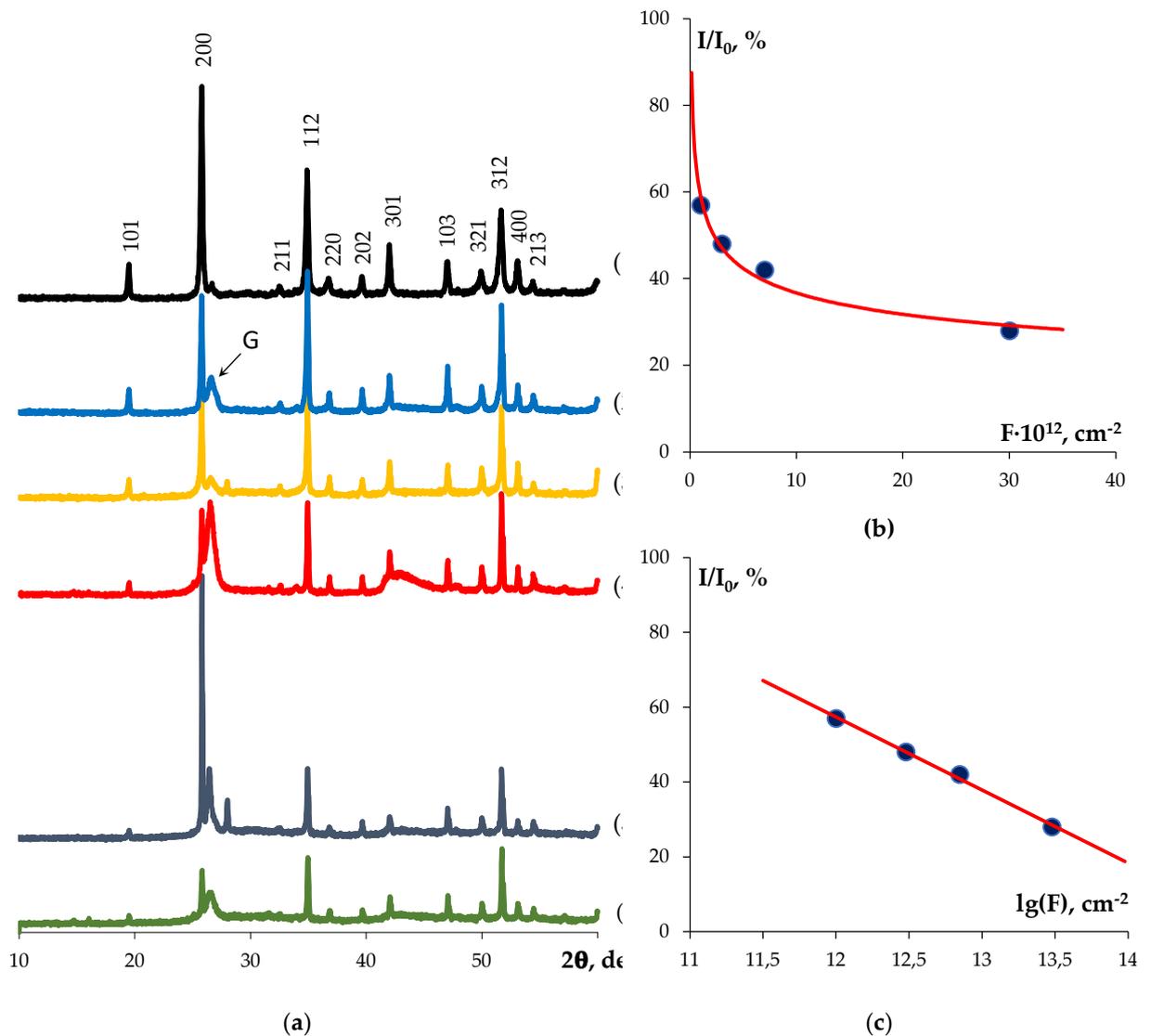

**Figure 6.** XRD data for the $YPO_4$:Gd ceramics after irradiation (**a**) with $Xe^{+26}$ ions. Fluences, cm$^{-2}$: (1) 0, (2) $1\times10^{12}$, (3) $3\times10^{12}$, (4) $7\times10^{12}$, (5) $1\times10^{13}$, (6) $3\times10^{13}$. Dependence of the relative intensity of the (200) XRD peak on the fluence in the linear (**b**) and logarithmic (**c**) axes. G – symbol marks the reflection from graphite.

The metallographic analysis of the side surface of the irradiated ceramic samples revealed the 10–20 µm thick surface layers having another color (Fig. 7).

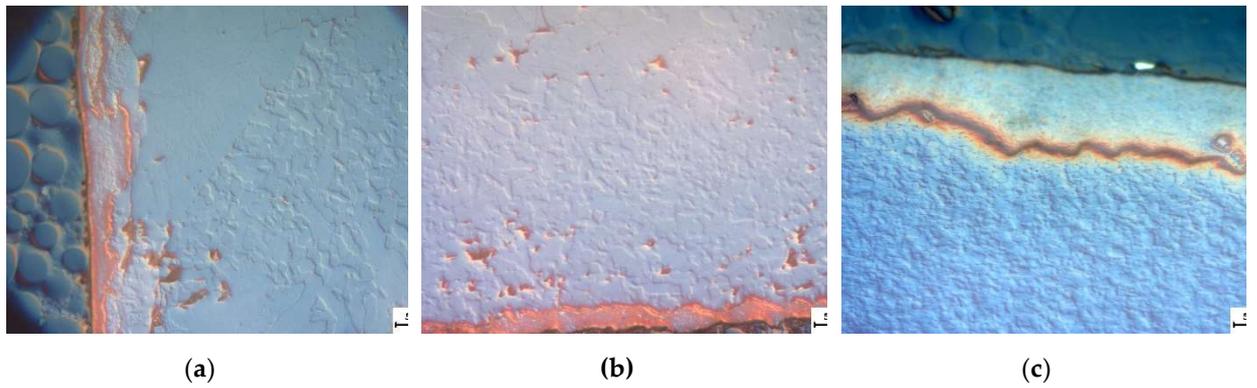

|     (a)     |     (b)     |     (c)     |

**Figure 7.** Side surfaces of the YPO$_4$:Gd ceramic samples irradiated with different fluences. F, cm$^{-2}$: (a) 1×10$^{12}$, (b) 3×10$^{12}$, (c) 3×10$^{13}$.

In our opinion, the change of the surface layer color originates from the carbonization of the ceramics owing to the interaction of the graphite mold and of the graphite paper with the ceramic sample surface. It is interesting to note that the carbonized layers were manifested more clear in the irradiated samples than in the initial ones. Note also that the carbonization of the surface layers in SPS was observed often for various metallic and ceramic materials [26,42–50]. Some authors suggest the carbonization of the materials in SPS originates from the low-temperature decomposition of the polymer used for binding of the graphite paper [49]. The authors of [49] suggested the gas-forming CO released from the graphite paper when heating to be the source of the "pollution" of the ceramic surface by carbon. In the course of rapid heating, the gas-phase CO penetrates inside the specimens through the open pores and then appears to be "locked" inside the pores upon achieving a high relative density. To suppress the intensive diffusion of carbon inside the ceramic surfaces, the sintering regimes with increased pressure [49], with step-wise change of the temperature and heating rate in SPS [50], application of molybdenum foil to protect the ceramics [44], etc. were proposed of the ceramics.

The recovery of the sample metamict phase after the irradiation by the maximum fluence of 3·10$^{13}$ cm$^{-2}$ was studied by sequential annealing at the temperatures increased stepwise from 200 to 700 °C for 3 hrs at each temperature. The XRD curves were measured after each step (Fig. 8). It follows from these data that the crystalline phase recovered already after annealing at 500 °C (total annealing time was 15 hrs). Further annealing at 600 °C promoted an increase in the intensity of the reflections. After annealing at 700 °C (total annealing time 18 h), the intensity of the diffraction peaks of the recovered sample reached ~ 80% of the initial one I$_0$.

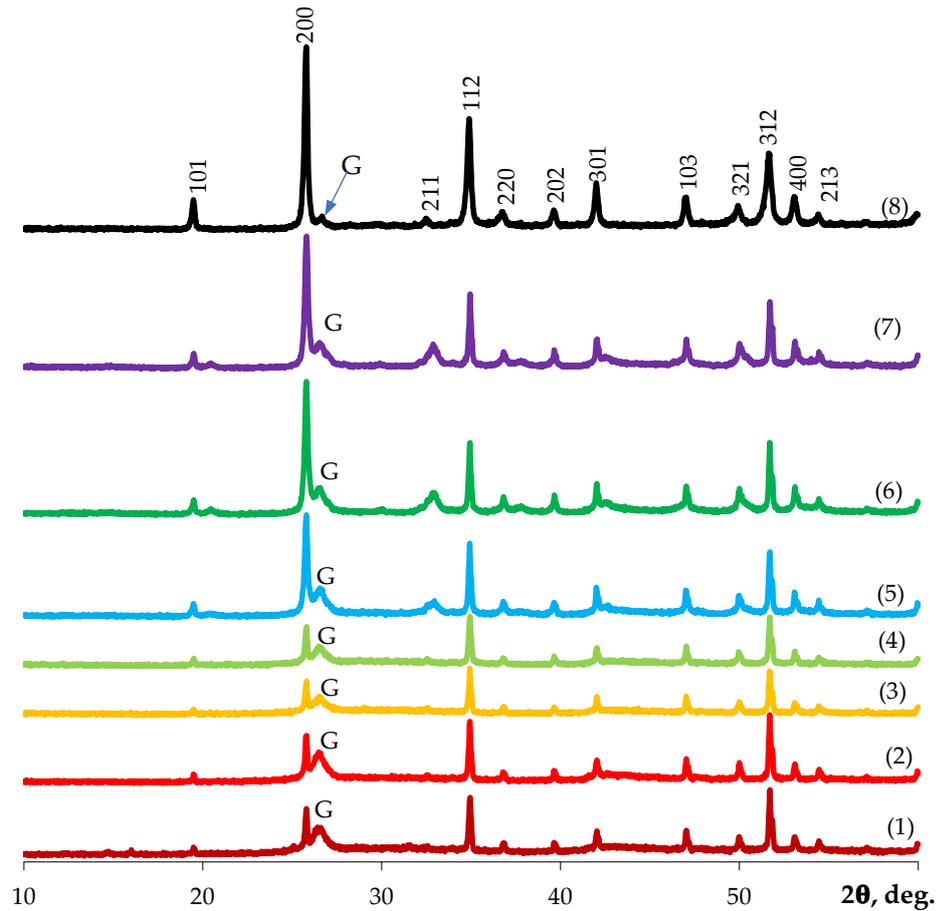

**Figure 8.** XRD data for the irradiated ceramic after irradiation (1) and after annealing at 200 °C (2), 300 °C (3), 400 °C (4), 500 °C (5), 600 °C (6), and 700 °C (7). (8) - XRD curve from the non-irradiated ceramic sample.

So far, the ceramics obtained are radiation-resistant in the conditions studied in the present work i.e. these can withstand the irradiation by the ion beams with high fluences without complete amorphization and can be restored to the high crystalline state by annealing.

To analyze the crystal structure of the surface layers of the ceramics in details, the GIXRD analysis of the initial and irradiated samples was performed. In the series of the GIXRD experiments, the 200 reflection of the $Y_{0.95}Gd_{0.05}PO_4$ phase was scanned for every sample. The incidence angle of the primary onto the sample varied in the range from 2° to 10° with the step of 1° that corresponded to the variation of the X-ray penetration depth inside the sample from 0.3 to 1.5 μm. The penetration depth was calculated according to the data from [51]. The calculated X-ray density of the material ($\rho_{th}$) was used in the calculations. It should be noted that the X-ray penetration depth in GIXRD analysis is relatively small (1.5 μm or less at the incidence angle 10°) and appears to be smaller than the carbonized layer depth estimated from the metallographic investigations. Typical results of the GIXRD analysis for the initial sample and for the ones irradiated with the $Xe^{+26}$ ions with the dose of $7\times10^{12}$ cm$^{-2}$ are presented in Fig. 9. The same experiments were carried out for all samples in the series.

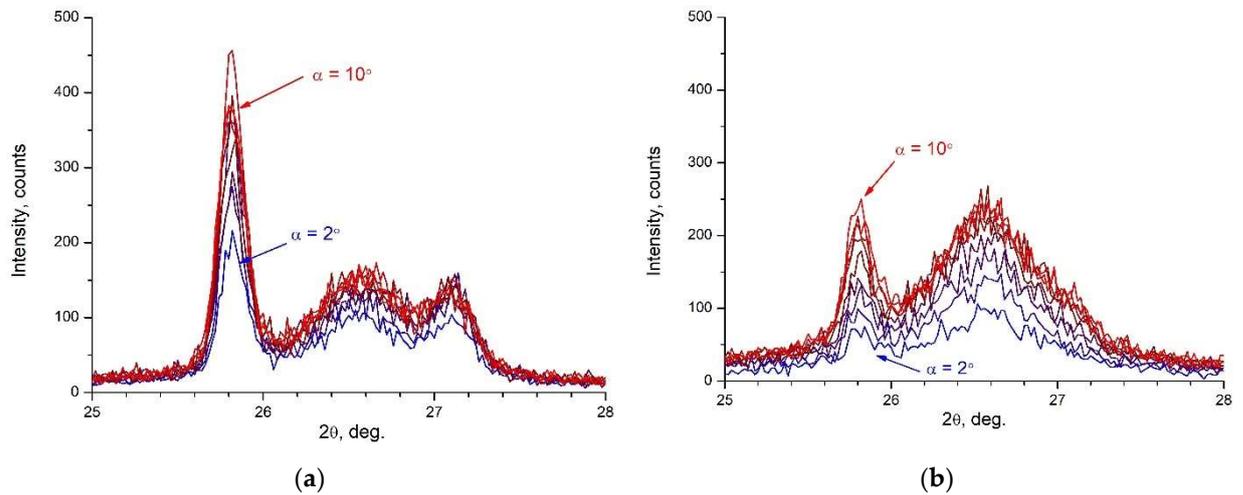

**Figure 9.** GIXRD analysis of the 200 reflections of the YPO$_4$:Gd ceramic samples: (**a**) initial state; (**b**) after irradiation with Xe$^{+26}$ ions with the dose of 7×10$^{12}$ cm$^{-2}$.

From the results presented in Fig. 9, one can see the presence of the main YPO$_4$:Gd phase as well as of the graphite one in the surface layer. In the initial sample, the diffraction peak from the graphite phase was split that may evidence a nonuniform distribution of this one inside the sample. In the irradiated sample, the splitting of the graphite peak disappeared that can originate from the recrystallization of the graphite phase in the equilibrium conditions after the ion irradiation.

To analyze the data of the GIXRD experiments, the dependencies of the integral intensity of the diffraction peaks of the YPO$_4$:Gd phase and of the graphite one on the X-ray incidence angle were plotted. Fig. 10a, these dependencies are plotted for (1) the initial YPO$_4$:Gd sample, (2) the sample after the irradiation with the Xe$^{+26}$ ions with the dose of 7×10$^{12}$ cm$^{-2}$, and (3) the sample after the irradiation with the dose of 3×10$^{13}$ cm$^{-2}$ followed by the recovery annealing. The triangles in the figures mark the dependencies of the intensity for the YPO$_4$:Gd phase, the hexagons – for the graphite one. Also, Fig. 10a shows the additional scale with the recalculation of the incidence angle into the information depth of the XRD analysis. One can see clearly the intensity of the YPO$_4$:Gd phase peak to decrease in the irradiated sample down to 30% (at the dose of 7×10$^{12}$ cm$^{-2}$) and to recover after annealing back to ~ 80% of the initial intensity. It evidences the formation of the amorphized surface layer, which recovers the crystallinity after annealing partly. The intensity of the graphite phase peak decreased after annealing that can be explained by the burnout of graphite during annealing.

To analyze the depth distributions of the phases, the dependencies in Fig. 10a were normalized in the intensities to the interval [0;1]. The dependencies of the normalized intensities for the diffraction peaks from the YPO$_4$:Gd phase (triangles) and from the graphite one (hexagons) on the primary beam incidence angle onto the sample are presented in Fig. 10b. One can see the intensity of the diffraction peak from the graphite phase to increase with increasing emission incidence angle onto the sample faster than the one from the Y$_{0.95}$Gd$_{0.05}$PO$_4$ phase both for the initial and irradiated ceramics. It evidences the graphite phase in these samples to be concentrated mainly in the subsurface layer. The irradiation resulted in the amorphization of the YPO$_4$:Gd phase in the whole range of the analysis depths but almost didn't affect the amorphization of the graphite phase. The annealing resulted in almost similar increase in the intensities both for the main YPO$_4$:Gd phase and for the graphite one with increasing penetration depth. It evidences more uniform depth distribution of graphite in the subsurface layer that can be explained by the burnout of carbon at the surface as well as by its diffusion inside the sample.

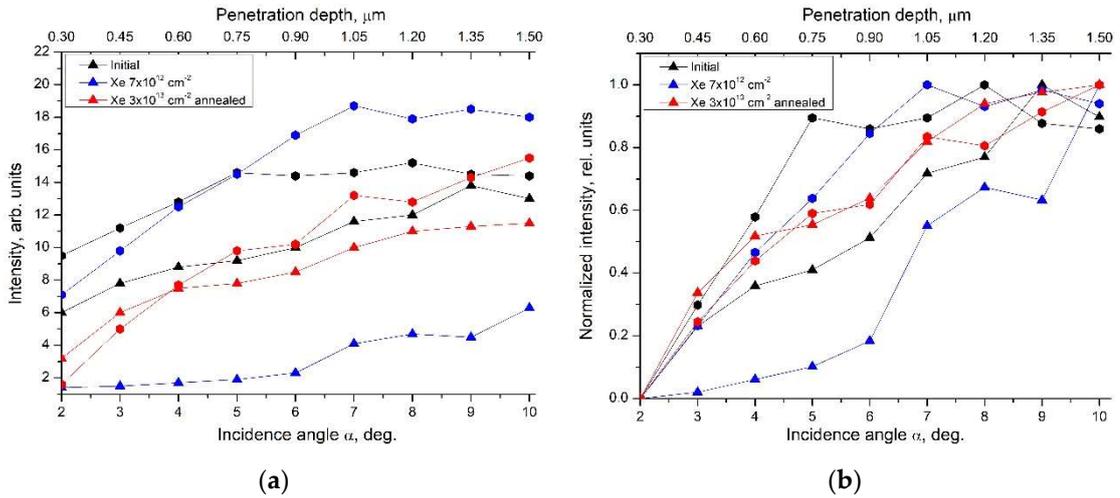

**Figure 10.** Analysis of the GIXRD results: (a) dependence of the intensity of XRD peaks from the YPO$_4$:Gd phase (triangles) and for the graphite one (circles) on the primary beam incidence angle onto the sample; (b) the same normalized in the intensity to the interval [0;1]. Black markers – initial sample, blue markers – after irradiation (F = 7×10$^{12}$ cm$^{-2}$), red markers – after irradiation and annealing 700 °C.

The calculations have shown the amorphization degree of the YPO$_4$:Gd phase in the subsurface layers to grow from 20% up to 70% with increasing fluence from 10$^{12}$ up to 3×10$^{13}$ cm$^{-2}$. After annealing, the sample with the maximum accumulated dose returned to the level of 20%.

Finally, let us determine the SPS activation energy for the YPO$_4$:Gd ceramics. To do so, let us use Yang-Cutler model describing the non-isothermic sintering of spherical particles at simultaneous diffusion inside the crystal lattice, grain boundary diffusion, and viscous flow of the material (creep) [52]. The applicability of Yang-Cutler model to the SPS of ceramics was proved in [33-38]. In [33-36,53-55], this model was applied to analyze the high-speed sintering of the fine-grained mineral-like ceramics for the HLW immobilization.

According to [52], the slope of the temperature dependence of the relative shrinkage ($\varepsilon$) in the $\ln(T \cdot \partial\varepsilon/\partial T) - T_m/T$ axes corresponds to the relative sintering activation energy $mQ_{s2}$ where $m$ is a coefficient depending on the dominating sintering mechanism ($m = 1/3$ for the grain boundary diffusion, $m = 1/2$ for the volume diffusion, and $m = 1$ for the viscous flow of the material (creep)). In the analysis of the results, the melting point $T_m$ was accepted to be 2273 K [12].

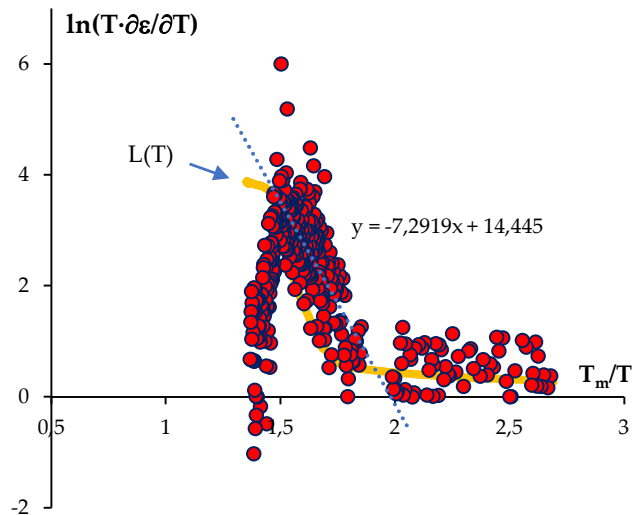

**Figure 11.** The dependence of $\ln(T \cdot \partial\varepsilon/\partial T) - T_m/T$ for YPO$_4$:Gd ceramic.

The dependencies $\ln(T \cdot \partial\varepsilon/\partial T) - T_m/T$ had usual three-stage character (see [52]). At the intensive compaction stage, the dependence $\ln(T \cdot \partial\varepsilon/\partial T) - T_m/T$ can be interpolated by a straight line with a good precision (Fig. 11). As one can see in Fig. 11, the effective activation energy $mQ_{s2}$ for the YPO$_4$:Gd ceramics was ~7.1 kT$_m$. For $m = 1/3$ typical for SPS of the fine-grained ceramics [33-36,53-55], the sintering activation energy $Q_{s2}$ for the YPO$_4$:Gd ceramics was 22 kT$_m$ (~ 210 kJ/mol). Unfortunately, there is no data on the activation energy of the grain boundary diffusion in the YPO$_4$:Gd ceramics at present. At the same time, it is worth noting that this value of $Q_{s2}$ is close to the activation energy of the grain boundary diffusion for many ceramic materials [33-38,40,53-55]. It allows suggesting the accelerated sintering of the YPO$_4$:Gd powders in SPS to originate from the intensive grain boundary diffusion at ~1200-1400 °C. The grain growth observed in the ceramics (Fig. 5), the rate of which also depends on the diffusion the intensity of the grain boundary diffusion supports this suggestion indirectly.

## 4. Conclusions

1. The phosphate powders Y$_{0.95}$Gd$_{0.05}$PO$_4$ where Gd acted as a simulator of Cm were obtained by sol-gel method. The ceramic samples with a high relative density (> 99%) were obtained by Spark Plasma Sintering (SPS). The sintering temperature was 1400 °C, the whole sintering process duration was ~18 min. The activation energy of SPS for the YPO$_4$:Gs fine-grained ceramic was ~22 kT$_m$ (~210 kJ/mol).

2. The ceramic samples demonstrated a high resistance to irradiation with Xe ions with the energy of 167 MeV. At the maximum irradiation fluence of 3·10$^{13}$ cm$^{-2}$, the surface layers of the ceramic samples retained the crystallinity partially. The calculated value of the fluence leading to the complete amorphization of the surface layers was (9.2 ± 0.1)·10$^{14}$ cm$^{-2}$. After annealing at 500 °C, the metamict phase recovered. After heating at 700 °C, the degree of recovery reached ~ 80%.

3. GIXRD experiments revealed the presence of the graphite phase concentrated mainly near the surfaces of the ceramic samples. The irradiation with the high energy ions resulted in the amorphization of the YPO$_4$:Gd phase in the subsurface layers and affected the crystallinity of the graphite phase weakly. The increasing of the irradiation dose resulted in the increase in the amorphization degree of YPO$_4$:Gd from 20% up to 70%. Subsequent annealing of the samples resulted in the decrease in the amorphization degree down to the level of 20% as well as in probable burnout and diffusion of carbon inside the samples that was manifested in more uniform depth distribution of the graphite phase in the annealed sample.

**Author Contributions:** Conceptualization, A.O., D.M. and E.P.; methodology, D.M., E.P. and A.O.; formal analysis, D.M., E.P., P.Y. V.C. and A.O.; investigation, D.M., E.P., P.Y., N.S., M.B., O.B., V.S., A.I., N.T.; resources, A.O., V.C., A.N.; data curation, A.N., A.O., V.C.; writing—original draft preparation, D.M., E.P., P.Y.; writing—review and editing, A.N., A.O.; visualization, A.N.; supervision, A.O.; project administration, A.N.; funding acquisition, A.N. All authors have read and agreed to the published version of the manuscript.

**Funding:** The study was funded by RFBR and ROSATOM (Grant No. 20-21-00145). The XRD investigations of the specimens after ion irradiation were carried out in Laboratory of Diagnostics of Radiation Defects in Solid State Nanostructures at Institute for Physics of Microstructures RAS (IPM RAS) with the financial support of the Ministry of Science and Higher Education of the Russian Federation (Grant No. 0030-2021-0030). The TEM study of the powders was carried out on the equipment of the Center Collective Use "Materials Science and Metallurgy" (NUST "MISIS") with the financial support of the Ministry of Science and Higher Education of the Russian Federation (Grant No. 075-15-2021-696).

**Institutional Review Board Statement:** Not applicable.

**Informed Consent Statement:** Not applicable.

**Data Availability Statement:** Not applicable.

**Conflicts of Interest:** The authors declare no conflict of interest.